\documentclass[fleqn,usenatbib]{hofmann}

\usepackage{newtxtext,newtxmath}

\usepackage{float}
\restylefloat{figure}
\restylefloat{table}
\usepackage{tabularx}
\usepackage[toc,page]{appendix}

\usepackage[paper=a4paper,left=17mm,right=17mm,top=20mm,bottom=20mm]{geometry}

\usepackage{caption}
\usepackage{setspace}

\usepackage{natbib}

\usepackage{color,soul}

\usepackage{hyperref}
\hypersetup{
    colorlinks=true,
    linkcolor=ultramarine3,
    filecolor=ultramarine3,      
    urlcolor=ultramarine3,
    pdftitle=Frequency-redshift relation of the Cosmic Microwave Background,
    pdfpagemode=FullScreen,
    citecolor=ultramarine3
    }

\usepackage{xcolor}
\definecolor{ultramarine}{RGB}{0,32,96}
\definecolor{ultramarine2}{RGB}{0,68,204}
\definecolor{ultramarine3}{RGB}{0,0,180}

\usepackage[T1]{fontenc}

\DeclareRobustCommand{\VAN}[3]{#2}
\let\VANthebibliography\thebibliography
\def\thebibliography{\DeclareRobustCommand{\VAN}[3]{##3}\VANthebibliography}

\usepackage{graphicx,subfigure}	
\usepackage{amsmath}	
\usepackage{amsmath,array,graphicx}
\usepackage{kantlipsum}
\usepackage{hyperref}

\newcommand{\eqb}{\begin{equation}}
\newcommand{\eqe}{\end{equation}}
\newcommand{\dmb}{\begin{displaymath}}
\newcommand{\dme}{\end{displaymath}}

\newcommand{\eab}{\begin{eqnarray}}
\newcommand{\eae}{\end{eqnarray}}

\newcommand{\be}{\begin{equation}}
\newcommand{\ee}{\end{equation}}

\title[Frequency-redshift relation]{Frequency-redshift relation of the Cosmic Microwave Background}

\author[Hofmann \& Meinert]{Ralf Hofmann$^{1}$\thanks{E-mail: R.Hofmann@ThPhys.Uni-Heidelberg.de}\href{https://orcid.org/0000-0001-6365-0631}{\hspace{0.1mm}\includegraphics[scale=0.06]{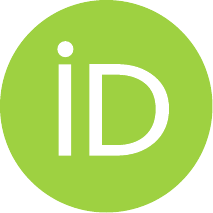}} and Janning Meinert$^{2,1}$\thanks{E-mail: J.Meinert@ThPhys.Uni-Heidelberg.de}\href{https://orcid.org/0000-0001-7582-3456}{\hspace{0.1mm}\includegraphics[scale=0.06]{orcid.pdf}}\\
\vspace{1.3mm}\\ 
$^{1}$Institut f\"ur Theoretische Physik, Universit\"at Heidelberg Philosophenweg 12, D-69120 Heidelberg, Germany\vspace{1mm}
\\ 
$^{2}$Department of Physics, Bergische Universit\"at Wuppertal, Gaußstraße 20, D-42119 Wuppertal, Germany\vspace{1mm}
\\
}

\begin{document}
\label{firstpage}
\pagerange{\pageref{firstpage}--\pageref{lastpage}}
\maketitle


\begin{abstract}
We point out that a modified temperature--redshift relation ($T$-$z$ relation) of the cosmic microwave background (CMB) can not be deduced by any observational method that appeals to an a priori thermalisation to the CMB temperature $T$ of the excited states in a probe environment of independently determined redshift $z$. For example, this applies to quasar-light absorption by a damped Lyman-alpha system due to atomic as well as ionic fine-splitting transitions or molecular rotational bands. Similarly, the thermal Sunyaev-Zel'dovich (thSZ) effect cannot be used to extract the CMB's $T$-$z$ relation. This is because the relative line strengths between ground and excited states in the former and the CMB spectral distortion in the latter case both depend, apart from environment specific normalisations, solely on the dimensionless spectral variable $x=\frac{h\nu}{k_B T}$. Since literature on extractions of the CMB's $T$-$z$ relation always assumes (i) $\nu(z)=(1+z)\nu(z=0)$ where $\nu(z=0)$ is the observed frequency in the heliocentric rest frame, the finding (ii) $T(z)=(1+z)T(z=0)$ just confirms the expected {\sl blackbody nature} of the interacting CMB at $z>0$. In contrast to emission of isolated, directed radiation, whose frequency-redshift relation ($\nu$-$z$ relation) {\sl is} subject to (i), a non-conventional $\nu$-$z$ relation $\nu(z)=f(z)\nu(z=0)$ of pure, isotropic blackbody radiation, subject to adiabatically slow cosmic expansion, necessarily has to follow that of the $T$-$z$ relation $T(z)=f(z)T(z=0)$ and vice versa. In general, the function $f(z)$ is determined by energy conservation of the CMB fluid in a Friedmann-Lemaitre-Robertson-Walker universe. If the pure CMB is subject to an SU(2) rather than a U(1) gauge principle, then $f(z)= \left({1/4}\right)^{1/3}(1+z)$ for $z\gg 1$, and $f(z)$ is nonlinear for $z\sim 1$. 
\end{abstract}




\section{Introduction}

Angular correlations between directionally dependent temperature and polarisation fluctuations of the 
Cosmic Microwave Background (CMB) radiation \citep{1994ApJ...420..439M} are important probes for the extraction of 
cosmological parameters \citep{Aghanim:2018eyx}. Since the observed angular correlations {are mainly} 
influenced by curvature-induced dark-matter potentials, which in turn cause acoustic oscillations of the baryon--electron--photon plasma prior to recombination, these parameters depend on high-$z$ physics when extracted from CMB data. 
{Therefore, they are very sensitive to the temperature redshift relation}
($T$-$z$ relation) which is assumed in expressing the CMB's energy density $\rho(T)$ 
in terms of $z$. \\

If the CMB is subject to a quantum U(1) gauge theory, then, according to the Stefan--Boltzmann law and energy conservation in a Friedmann--Lemaitre--Robertson--Walker (FLRW) universe, the $T$-$z$ relation is $T(z)/T(z=0)=z+1$, where $T(z=0)$ is today's CMB temperature $T(z=0)=2.726 $K \citep{1994ApJ...420..439M}.

Such a U(1) $T$-$z$ relation is identical to the frequency-redshift relation ($\nu$-$z$ relation) 
$\nu(z)/\nu(z=0)=z+1$ describing electromagnetic waves emitted by compact astrophysical objects 
and travelling through an expanding FLRW universe towards the observer \citep{Weinberg:1972kfs}.\\


The thermodynamics underlying the CMB {and} the thermodynamics of a dense gas of absorber--emitter particles may be richer than they appear, such that the two situations need to be distinguished.
While the CMB can be represented by a photon gas within its bulk, absorber--emitter particles thermalise via electromagnetic waves whose emissions and absorptions are enabled by electronic transitions. 
Therefore, the conventional $T$-$z$ relation may not hold 
universally but, depending on how the above two extreme situations are mixed, is modified as 
$T(z)/T(z=0)=f(z)$, where the function $f(z)$ is specific to the generalising theory. Thermodynamics then immediately implies that also the $\nu$-$z$ relation 
is also of the form $\nu(z)/\nu(z=0)=f(z)$. Such is the case, for example, 
if the thermal photons (and low-frequency waves) of the CMB are identified with the Cartan modes of a single thermal SU(2) Yang--Mills theory, SU(2)$_{\rm CMB}$,  
in the deconfining phase \citep{Hofmann:2016cqh,Hahn:2017yei}.   
These modes interact only feebly within a small range of low 
temperatures and frequencies with the two off-Cartan quasiparticle vector modes. The fact that all gauge modes, massless and massive, are excitations of one and the same thermal ground-state adds additional $T$ dependent energy density to that of thermal fluctuations: the ground-state energy density rises linearly in $T$ in contrast to 
the rapidly attained Stefan-Boltzmann law $\propto T^4$ associated with thermal fluctuations \citep{Hofmann:2016cqh}.

If the CMB as a bulk thermal photon gas is indeed subject to SU(2)$_{\rm CMB}$ thermodynamics (single Yang--Mills theory in its deconfining phase), from $T = 7.99\,$keV (or $T=1.09\times 10^8$ K) to $T = 2.3\times 10^{-4}$ eV (or $T=2.726$ K) then a number of 
implications arise, see \cite{https://doi.org/10.1002/andp.202200517,Hahn:2018dih,Szopa:2007wy,Hofmann:2013rna} 
for the CMB large-angle anomalies, \cite{Hahn:2017hjt,Hahn:2018dih} for the modified high-$z$ cosmological model
implied by a modified $T$-$z$ relation \cite{Hahn:2017yei}, \cite{MeinertHofmann2021} for dark-sector physics, and \cite{Hofmann:2014lka} for neutrinos.\\

The purpose of the present paper is to point out that past observational extractions of the 
CMB's $T$-$z$ relation from background-light absorbing systems, which are assumed to thermalise with the CMB in a conventional way, are bound to 
extract the standard U(1) $T$-$z$ relation if participating frequencies (observed absorption lines) are blueshifted accordingly. 
This is also true of the observation of spectral CMB distortions 
inflicted by its photons scattering off hot electrons belonging to X-ray clusters along the line of sight, i.e., the thermal Sunyaev-Zel'dovich effect (thSZ). In Section\,\ref{OS}, we discuss these two observational approaches in 
more detail. First, we analyse the extraction of $T(z)$ from absorption lines within 
the continuous spectrum of a background source caused by a cloud in its line of sight, 
which is assumed to be thermalised with the CMB. Second, we discuss the distortions of the CMB spectrum according to the thSZ effect. 
In the former case, the frequency of the absorption line $\nu(z)=(z+1)\nu(z=0)$, which is assumed to coincide with the exciting CMB frequency, 
is used to extract a temperature $T(z)$. 
Note that in this case $T(z)$ coincides with the present CMB's temperature $T(z=0)$ only if it is redshifted as $T(z)/(z+1)=T(z=0)$.

In other words, ignoring the value of a known transition frequency $\nu^*(z)$ of the system in using a different $\nu$-$z$ relation for $\nu(z)$, 
$\nu(z)=f(z)\nu(z=0)$, the extracted CMB temperature would only have redshifted to its present 
value under the use of $T(z)/f(z)=T(z=0)$. Therefore, it appears that in a given absorber system, 
interaction with the CMB occurs by a local shift of the CMB frequency $\nu(z)$ and temperature $T(z)$
to the line frequency $\nu^*(z)=(z+1)\nu(z=0)$ and cloud temperature $T^*(z)=(z+1)T(z=0)$ 
of the absorbing molecules. For deconfining SU(2) Yang--Mills thermodynamics, $\nu(z) \rightarrow \nu^*(z)$ is an upward shift (see Section\,\ref{YM}).\\


The thermalisation within a photon gas far away from any charges is different from the thermalisation within absorber clouds. This is because the degrees of freedom invoked are not the same.

As a result, $T^*(z)=(z+1)T(z=0)$ is interpreted as the 
CMB's $T$-$z$ relation, while it is actually $T(z)=f(z)T(z=0)$. In exploiting the thSZ effect for the CMB's $T$-$z$ relation extractions, 
we observe a similar situation. In Section\,\ref{YM}, we review ~\cite{Hofmann:2016cqh,Hahn:2017yei} how a modified $T$-$z$ relation (and then the $\nu$-$z$ relation) 
arises if the CMB is subject to deconfining SU(2) rather than U(1) quantum thermodynamics and how the Yang--Mills scale of such an SU(2) model, in the following referred to as SU(2)$_{\rm CMB}$, is fixed by observation. To archive this, the CMB radio excess in line temperature, see, e.g., \cite{Fixsen:2009xn,Dowell_2018}, is interpreted as an effect due to the transition between deconfining and preconfining SU(2) Yang--Mills thermodynamics. We also discuss a number of alternative explanations of this effect.\\

Moreover, we discuss how another SU(2) model, SU(2)$_{\rm e}$ \citep{2017Entrp..19..575H,Hofmann:2020yhn,2016tqym.book.....H}, whose two stable solitonic 
excitations in the confining phase represent 
the first-family lepton doublet, mixes with SU(2)$_{\rm CMB}$.
Such a mixing depends,
up to temperatures of $\sim 7.99$ keV, on the degree of thermalisation prevailing in a local environment of electromagnetically interacting electronic charges within a certain range of frequencies and charge densities. 
The aforementioned upward shift in the CMB frequency $\nu(z)$ to absorption line frequency $\nu^*(z)$, accompanied by a shift in the CMB temperature $T(z)$ to cloud temperature $T^*(z)$, would then be a consequence of an incoherent mixture 
of the Cartan modes of SU(2)$_{\rm CMB}$ (thermal photonic fluctuation) with those of SU(2)$_{\rm e}$ (thermalised electromagnetic waves) 
when moving from empty space to the interior of the cloud. 
Finally, in Section\,\ref{SC}, we summarise the results of this paper; mention an 
observational signature which is sensitive to the CMB's $T$-$z$ relation, the spectrum of ultra-high energy cosmic rays (UHECRs); and briefly discuss implications for Big Bang nucleosynthesis.\\

From now on, we work in natural units $c=k_B=\hbar=1$, where $c$ denotes the speed of light in vacuum, $k_B$ is Boltzmann's constant, and 
$\hbar$ refers to the reduced quantum of action.

\section{Observational \texorpdfstring{$T$-$\MakeLowercase{z}$}{T-z} relation extractions from a prescribed \texorpdfstring{$\nu$-$\MakeLowercase{z}$}{v-z}
relation\label{OS}}

In this section, we discuss two principle probes used in the literature to extract the redshift 
dependence of the CMB temperature $T(z)$ up to $z\sim 6.34$, see \cite{2022Natur.602...58R} for a useful compilation, and how these 
extractions are prejudiced by an assumed $\nu$-$z$ relation of CMB frequencies.\\ 

The first class of probes comprises absorbing clouds of known redshifts, e.g., parts of damped Lyman-$\alpha$ systems, in the line of sight of a distant 
quasar or a bright galaxy. Here, the assumed thermalisation with the CMB populates the fine-structure 
levels of the ground-states of certain atoms or ions \citep{1994Natur.371...43S,1997ApJG,2000Natur.408..931S,2002A&A...381L..64M,2022Natur.602...58R} or excites 
rotational levels of certain molecules \citep{2000Natur.408..931S,2011A&A...526L...7N,2017A&A...597A..82N} whose 
population ratios can be obtained from the respective absorption-line profiles within the broad background spectra.\\ 

Modelling environmentally dependent contributions to level populations, such as particle collisions or pumping by UV radiation, the relative level populations yield upper-limit estimates of $T(z)$ at the redshift of the cloud.  
The limitations of this method are discussed in \cite{2017ApJ...847...65M}.
Note that in \cite{2013A&A...551A.109M}, a solution of the rotational excitations of various molecular species could be provided directly from their spectra.\\

The second class of probes refers to the observation of the CMB spectrum within certain frequency bands 
along the lines of sight of X-ray clusters of known redshifts. A characteristic spectral distortion, known as the thermal Sunyaev-Zel'dovich effect (thSZ) \citep{1969Ap&SS...4..301Z,1972CoASP...4..173S} and caused by inverse Compton scattering of CMB photons off free, thermal electrons of these clusters is exploited to estimate $T(z)$.  

\subsection{Absorber clouds in the line of sight of a quasar or a bright galaxy}

Estimates of $T(z)$ using the relative populations due to the excitation of atomic (ionic) fine-structure transitions and 
molecular rotation levels by the CMB have a long history, see \cite{1968ApJ...152..701B} 
for the theoretical basis and \citep{2011A&A...526L...7N,2017A&A...597A..82N,1994Natur.371...43S,2000Natur.408..931S,2013A&A...551A.109M,2022Natur.602...58R} for applications. Since sources of level 
excitations other than the CMB (e.g., collisions, UV pumping) have to be modelled for 
a given absorber, the extracted $T(z)$ is usually seen as an upper bound on the true CMB temperature.\\   

If the CMB is assumed to be the sole source of level population then the 
extraction of $T(z)$ is facilitated in terms of the column density of the 
absorber species, depending on the measured line strength in the continuous spectrum of the background source, the transition frequencies, the temperature at which levels are thermalised, and the integrated opacity of the line. Apparently, this method is validated by the measurement in \cite{1993ApJ...413L..67R} of the CMB temperature $T(z=0)=(2.726^{+0.023}_{-0.031})\,$K in our Galaxy, analysing CN rotational 
transitions in five diffuse interstellar clouds. The value extracted in this way, $T(z\,=\,0)$, agrees well with the spectral CMB fit by COBE \citep{1994ApJ...420..439M} of $T(z=0)=(2.726\pm 0.010)\,$K. 
So what about $z>0$?\\

In \cite{2013A&A...551A.109M} a molecular rich cloud within a 
spiral galaxy at $z=0.89$ was observed towards the radio-loud, gravitationally lensed blazar 
PKS 1830-211 at redshift $z=2.5$. Within the cloud, the rotational temperature $T_{\rm rot}$ is defined via
\eqb
\label{defrotT}
\frac{n_u}{n_l}=\frac{g_u}{g_l}\,\exp\left(-\frac{2\pi\nu^*(z)}{T_{\rm rot}}\right)\,,
\eqe
where $n_u$ ($g_u$) and $n_l$ ($g_l$) are the populations (degeneracies) of the upper and lower level, respectively, 
and $\nu^*(z)$ denotes the transition frequency in the cloud's rest frame. For the rotational excitations 
of ten molecules, $T_{\rm rot}$ was interpreted to universally 
represent $T(z=0.89)$ because the molecular gas was estimated 
to be sub-thermally excited (rotational levels 
solely {\sl radiatively} coupled to the CMB, negligible impact of 
collisions and the local radiation field). Observations were performed within 3 wavelength 
bands at around $\lambda=2,3,7$\,mm using two different instruments. In one (simplified) approach, the extraction of $T_{\rm rot}$ from the 
two transitions in each molecular species was performed by pinning down the intersection 
of the two column densities $N_{\rm LTE}$ depending on $T_{\rm rot}$. 
For a given transition, $N_{\rm LTE}$ is defined as
\eqb\label{NLTE}
N_{\rm LTE}=\frac{3}{4\pi^2\mu^2 S_{ul}}Q(T_{\rm rot})\frac{\exp\left(\frac{E_l}{T_{\rm rot}}\right)}
{1-\exp\left(-\frac{2\pi\nu^*(z)}{T_{\rm rot}}\right)}\int\tau dv\,,
\eqe
where $E_l$ is the energy of the lower level, $Q(T_{\rm rot})$ the partition function including 
all rotational excitations, $\mu$ the dipole moment, $S_{ul}$ the observed line strength, and 
$\int\tau dv$ the integrated (observed) opacity of the line. Across the absorption lines of all molecular species considered, this approach 
produces values of $T_{\rm rot}$, which are quite consistent with the expectation $T(z=0.89)=(1+0.89)T(z=0)=5.14\,$K.\\


\noindent Setting $T_{\rm rot}=T^*(z)$ in Eq.\,(\ref{NLTE}) and using $\nu^*(z)=(1+z)\nu(z=0)$ is only consistent with the participating CMB photons being distributed according to a blackbody spectrum if $T^*(z)$ also redshifts as $T^*(z)=(1+z)T(z=0)$; $\nu(z=0)$ denotes the observed frequency of the transition in the heliocentric restframe. Therefore, this is an in-built feature of the model even though the
CMB may, in reality, exhibit a different $T$-$z$ relation (and then also $\nu$-$z$ relation)\footnote{One may think of the true CMB temperature (which would be lower in SU(2)$_{\rm CMB}$) and participating 
CMB frequency being elevated by the same factor to $T^*(z)$ and $\nu^*(z)$ of a rotational excitation, respectively, by an incoherent mixing of a Cartan mode in SU(2)$_{\rm CMB}$ and a Cartan mode of SU(2)$_{\rm e}$ as the observer moves from empty space outside the cloud towards its interior.}.\\

The situation is similar for the observation of atomic / ionic fine-structure transitions in 
absorbers at $z>0$. Also, here, the very assumption of these excitations thermalising with the 
CMB ties the extracted $T$-$z$ relation to the $\nu$-$z$ relation used in converting 
observed (heliocentric) frequencies to transition frequencies 
in an absorber's restframe: The proper use of $f(z)=1+z$ for absorption lines produces a 
higher cloud temperature $T^*(z)$ than CMB temperature $T(z)$ if the latter is 
assumed to be described by an unmixed SU(2) model, see Section\,\ref{YM}.\\
%

In Section\,\ref{APE}, we will discuss in more detail a degree-of-thermalisation dependent mixing of Cartan excitations 
in two SU(2) gauge groups explaining why directed radiation, as issued by the background source and observed in a 
spectrally resolved way after having passed the absorber, obeys a conventional $\nu$-$z$ relation 
while the $\nu$-$z$ relation of CMB photons necessarily follows that of the $T$-$z$ relation, which may well be 
unconventional \citep{Hahn:2017yei}.

\subsection{The thermal Sunyaev-Zel'dovich effect\label{SZ}}

The thermal Sunyaev-Zel'dovich effect (thSZ) is a distortion of the blackbody shape of the CMB spectrum that is induced by inverse Compton scattering of CMB photons off thermalised electrons in the X-ray plasmas of a given cluster of galaxies \citep{1969Ap&SS...4..301Z,1972CoASP...4..173S}. Neglecting contributions from the weakly relativistic high-end part of 
the electrons' velocity distribution, the thSZ effect can be 
expressed in terms of a frequency dependent (line-temperature) shift 
$\Delta T$ with respect to CMB baseline temperature $T$ at the cluster's redshift $z$ as \citep{2014A&A...561A.143H}

\eab
\label{thSZ}
&&\frac{\Delta T}{T}(x,\vec{n})=\nonumber\\ 
&&\left[\frac{\sigma_T}{m_e}\int ds\,
n_e(s,\vec{n})\cdot T_e(s,\vec{n})\right]\cdot\left[x\coth\left(\frac{x}{2}\right)-4\right]\,.
\eae
Here $m_e$ and $\sigma_T$ refer to the mass of the electron and the 
Thomson cross-section, respectively. Both the electron temperature $T_e$ and the electron 
number density $n_e$ depend on the proper distance parameter $s$ along the direction $\vec{n}$ of 
the line of sight under which CMB photons interacting with a given X-ray cluster are observed. 
The dimensionless variable $x$ is defined as $x\equiv\frac{2\pi \nu}{T}$. As Eq.\,(\ref{thSZ}) 
indicates, the thSZ effect factorises into an environmental part, determined 
by the thermodynamics of the X-ray cluster at redshift $z$ and dubbed {\sl thSZ flux}, and into 
a part which solely depends on $x$. We note that the zero $x_0$ 
of the second factor is 
\eqb
\label{dimlesstoday}
x_0\sim 3.83=\frac{\nu_0}{T(z=0)}\,,
\eqe
where $T(z=0)=2.726\,$K \citep{1994ApJ...420..439M} denotes the CMB temperature 
today. As a consequence, the thSZ effect predicts $\nu_0\sim 217\,$GHz. 
The $z$ dependence of $T$ already can be extracted by focusing on the frequency 
$\nu_0$ at which $\frac{\Delta T}{T}(x,\vec{n})$ vanishes. By virtue of Eq.\,(\ref{dimlesstoday}) a blueshift of $\nu_0$ according to the $\nu$-$z$ relation  
\eqb
\label{blueshift_fr}
\nu^*_0(z)=f(z)\nu_0
\eqe
then yields the $T$-$z$ relation 
\eqb
\label{TzthSZ0}
T(z)=\frac{f(z)\nu_0}{x_0}=f(z)T(z=0)\,.
\eqe
Therefore, whatever the assumption on $f(z)$ in the $\nu$-$z$ relation of Eq.\,(\ref{blueshift_fr}) is, this assumption necessarily transfers to the $T$-$z$ relation of Eq.\,(\ref{TzthSZ0}) if the intensity of the unperturbed CMB at any 
redshift $z>0$ is to possess a {\sl blackbody frequency distribution
\footnote{To a very good approximation the spectral intensity $I(\nu)$ of today's CMB is given as 
$I_{z=0}(\nu)d\nu=16\pi^2\frac{\nu^3}{\exp\left(\frac{2\pi\nu}{T(z=0)}\right)-1}d\nu$ \citep{1994ApJ...420..439M}. If we assume a $T$-$z$ relation of 
$T(z=0)=\frac{1}{f(z)}T(z)$ and a $\nu$-$z$ relation of $\nu(z=0)\equiv \frac{1}{g(z)}\nu^\prime$ with $f(z)\not=g(z)$, 
then the Stefan-Boltzmann law would still have redshifted according to the $T$-$z$ relation: 
$\int d\nu I_{z=0}(\nu)=\frac{\pi^2}{15}T^4(z=0)=
\frac{\pi^2}{15}\left(\frac{T(z)}{f(z)}\right)^4=
\left(\frac{1}{g(z)}\right)^4\int d\nu^\prime I_{z}(\nu^\prime)$. However, 
the maximum $\nu_{\rm max}=\frac{2.821}{2\pi}\,T(z=0)$ of the 
distribution $I_{z=0}(\nu)d\nu$ converts to a maximum 
$\nu^\prime_{\rm max}=\frac{2.821}{2\pi}\frac{g(z)}{f(z)}\,T(z)$ of 
the distribution $I_{z}(\nu^\prime)d\nu^\prime=
16\pi^2\frac{(\nu^\prime)^3}{\exp\left(\frac{f(z)}{g(z)}\frac{2\pi\nu^\prime}{T(z)}\right)-1}d\nu^\prime$. Thus, $I_{z}(\nu^\prime)$ no longer 
would be a blackbody spectrum.}}.\\

To suppress the statistical error in extractions of $T(z)$, a {\sl set} of frequency bands, centered at $\{\nu_i\}$, usually is invoked in fitting the 
modelled thSZ emission law to the observations with respect to X-ray clusters within a given redshift bin $\delta$. 
For example, in \cite{2014A&A...561A.143H} 
the Planck frequency bands at 100, 143, 217, 353, and 545\,GHz were used in multiple redshift bins, 
stacking of patches in a given redshift bin $\delta$ and frequency band centered at $\nu_i$ was performed, and the thSZ emission law 
was modelled by integrating Eq.\,(\ref{thSZ}) over bandpasses and by normalising it with 
the bandpass averaged calibrator emission law. Relevant fit parameters turned out to be the (stacked) 
thSZ flux $Y^\delta$, $T^\delta$, and the radio-source flux 
contamination $F^\delta_{\rm rad}$ which subsequently were estimated by a 
profile likelihood analysis. The crucial point here is that, in the modelling of the thSZ emission law within redshift bin 
$\delta$, a blueshifting of observation frequency $\nu_i$ to $\nu^*_i=f(z)\nu_i$ 
needs to be applied, implying again the $T$-$z$ relation 
\eqb
\label{TzthSZ}
T(z)=\frac{f(z)\nu_i}{x_i}=f(z)T(z=0)\,,
\eqe
where $x_i$ now is the solution to
\eqb
\label{otherfrequ}
\frac{F_i^\delta}{Y^\delta}=\coth\left(\frac{x_i}{2}\right)-4\,,
\eqe
and $F_i^\delta$ denotes the stacked, observed thSZ flux within redshift bin $\delta$. 
In \cite{2014A&A...561A.143H} the use of $\nu^*_i=(1+z)\nu_i$ thus 
necessarily leads to the conventional $T$-$z$ relation $T(z)=(1+z)T(z=0)$. To the best of the authors' knowledge, the use of $f(z)=1+z$ in the $\nu$-$z$ relation is, however, common to all extractions of $T(z)$ that appeal to the thSZ effect.\\ 

When the CMB gauge field represents the Cartan subalgebra of an SU(2) Yang--Mills theory, SU(2)$_{\rm CMB}$, it can be shown \citep{Hahn:2017yei} that $f(z)$ is different from $f(z)=1+z$ in the $T$-$z$ relation, see also Section\,\ref{TRRCMB}, and therefore also in the $\nu$-$z$ relation. 
This is because, in addition to thermal photons, the thermal ground-state in the 
deconfining phase of an SU(2) Yang--Mills theory is excited towards two vector modes subject to  
a temperature dependent mass. A feeble coupling of these two kinds of excitations leads 
to spectral distortions of CMB radiance deeply within the Rayleigh-Jeans 
regime \citep{https://doi.org/10.1002/andp.202200517} which is not targeted by Planck frequency bands. 
In Section\,\ref{YM} we review how this $T$-$z$ relation arises, and we discuss why an according non-conventional 
$\nu$-$z$ relation is to be expected from a thermalisation process which is dependent on the mixing angle between the Cartan modes of two SU(2) gauge models subject to disparate Yang--Mills scales.

\section{\texorpdfstring{$T$-$\MakeLowercase{z}$}{T-z} relation and \texorpdfstring{$\nu$-$\MakeLowercase{z}$}{v-z} relation in \texorpdfstring{SU(2)$_{\rm CMB}$}{SU(2)CMB}: Theoretical basis}\label{YM}

In this section, we discuss in detail why a 
thermal gas of electromagnetic disturbances (far away from any emitting surface on the scale of a 
typical inverse frequency $\nu^{-1}$) and with an isotropic and spatially homogeneous 
flux density of practically incoherent photons obeys a $T$-$z$ relation and an associated $\nu$-$z$ relation which are different 
from the $\nu$-$z$ relation of directed, (partially) coherent radiation. 

\subsection{\texorpdfstring{$T$-$z$}{T-z} relation in \texorpdfstring{SU(2)$_{\rm CMB}$}{SU(2)CMB}\label{TRRCMB}}


A pronounced distortion of the blackbody spectrum of radiance deeply within the Rayleigh-Jeans part was observed in \cite{Fixsen:2009xn, Dowell:2018mdb} and in references therein.
To explain this highly isotropic CMB radio excess at frequencies below 1\,GHz, we argued in \cite{Hofmann:2009yh} that the critical temperature $T(z=0)$ for the deconfining-preconfining phase transition of SU(2) Yang--Mills thermodynamics is very close to the present temperature of the CMB of $T(z=0)=2.726\,$K \citep{1994ApJ...420..439M}. 
This may seem to be a somewhat fine-tuned situation. 
However, the difference with an ordinary tuning of parameters by hand is that, 
the dual gauge coupling thermodynamically rises rapidly as the temperature drops into the preconfining phase. 
Since the Cartan mode's extracted thermal quasiparticle mass is 100 MHz, which is around three orders of magnitude smaller than the critical temperature $T_{\rm c}$, it follows that $T(z=0)$ needs to be very close to $T_{\rm c}$. However, due to a presently incomplete understanding 
of supercooling of the deconfining into the preconfining phase and associated tunnelling there is a tolerance of $\sim 10\,\%$ in this dynamic tuning of the two temperatures, which corresponds to about 1\,Gy of cosmic 
evolution \citep{2007EPJC...50..635G}. The exact assignment $T(z=0)=T_{\rm c}$, addressed further below,  
implies a Yang--Mills scale $\Lambda_{\rm CMB}=
1.064\times 10^{-4}\,$eV, and thus it is justified to 
refer to the SU(2) Yang--Mills model, whose deconfining thermodynamics is assumed to describe the CMB, as SU(2)$_{\rm CMB}$. 
We quote below a number of alternative approaches to explaining the CMB radio excess.

Let us now review \cite{Hahn:2017yei}, how the $T$-$z$ relation of 
deconfining SU(2)$_{\rm CMB}$ thermodynamics is derived from 
energy conservation in Friedmann-Lemaitre-Robertson-Walker 
(FLRW) universe of cosmological scale factor $a$, normalized such that today $a(T(z=0))=1$. One has 
\eqb
\label{enecons}
\frac{\mbox{d}\rho}{\mbox{d}a}=-\frac{3}{a}\left(\rho+P\right)\,,
\eqe
where $\rho$ and $P$ denote energy density and pressure 
of deconfining SU(2)$_{\rm CMB}$ thermodynamics, respectively. As usual, redshift 
$z$ and scale factor $a$ are related as $a^{-1}=z+1$. 
Eq.\,(\ref{enecons}) has the formal solution 
\begin{equation}
\label{formsol}
a=\exp\left(-\frac{1}{3}\int^{\rho(T)}_{\rho(T(z=0))}\frac{\mbox{d}\rho}{\rho+P(\rho)}\right) 
= \exp\left(-\frac{1}{3}\int^T_{T(z=0)}\mbox{d}T^\prime\,\frac{\frac{1}{T^\prime}\frac{\mbox{d}\rho}{\mbox{d}T^\prime}}{s(T^\prime)}\right) \,,
\end{equation}
where the entropy 
density $s$ is defined as 
\eqb
\label{entropydens}
s=\frac{\rho+P}{T}\,.
\eqe 
By virtue of the Legendre transformation 
\eqb
\label{LT}
\rho=T\frac{\mbox{d}P}{\mbox{d}T}-P\,,
\eqe
one has 
\eqb
\label{Tdiff}
\frac{1}{T}\frac{\mbox{d}\rho}{\mbox{d}T}=
\frac{\mbox{d}^2P}{\mbox{d}T^2}=
\frac{\mbox{d}s}{\mbox{d}T}\,.
\eqe 
Substituting Eq.\,(\ref{Tdiff}) into Eq.\,(\ref{formsol}) finally yields
\eab
\label{sol}
a &=& \frac{1}{z+1} = \exp\left(-\frac{1}{3}\int^T_{T(z=0)} \mbox{d}T^\prime \, \frac{\mbox{d}}{\mbox{d}T^\prime} 
\left[\log \frac{s(T^\prime)}{M^3} \right]\right)\nonumber\\ 
&=& \exp\left(-\frac13\log\frac{s(T)}{s(T(z=0))}\right)\,.
\eae 
Here, $M$ denotes an arbitrary mass scale. The formal solution
(\ref{sol}) is valid for any thermal and conserved 
fluid subject to expansion in an FLRW universe. 
If the function $s(T)$ is known, then (\ref{sol}) can be 
solved for the $T$-$z$ relation $T(z)$. 
Eqs.\,(\ref{entropydens}) and (\ref{sol}) exclude a ground-state 
dependence of the $T$-$z$ relation, since the equation of state for ground-state pressure $P^{\rm gs}$ and energy density $\rho^{\rm gs}$ is $P^{\rm gs}=-\rho^{\rm gs}$ \citep{2016tqym.book.....H}.\\

\begin{figure}
\centering
\includegraphics[width=\columnwidth]
{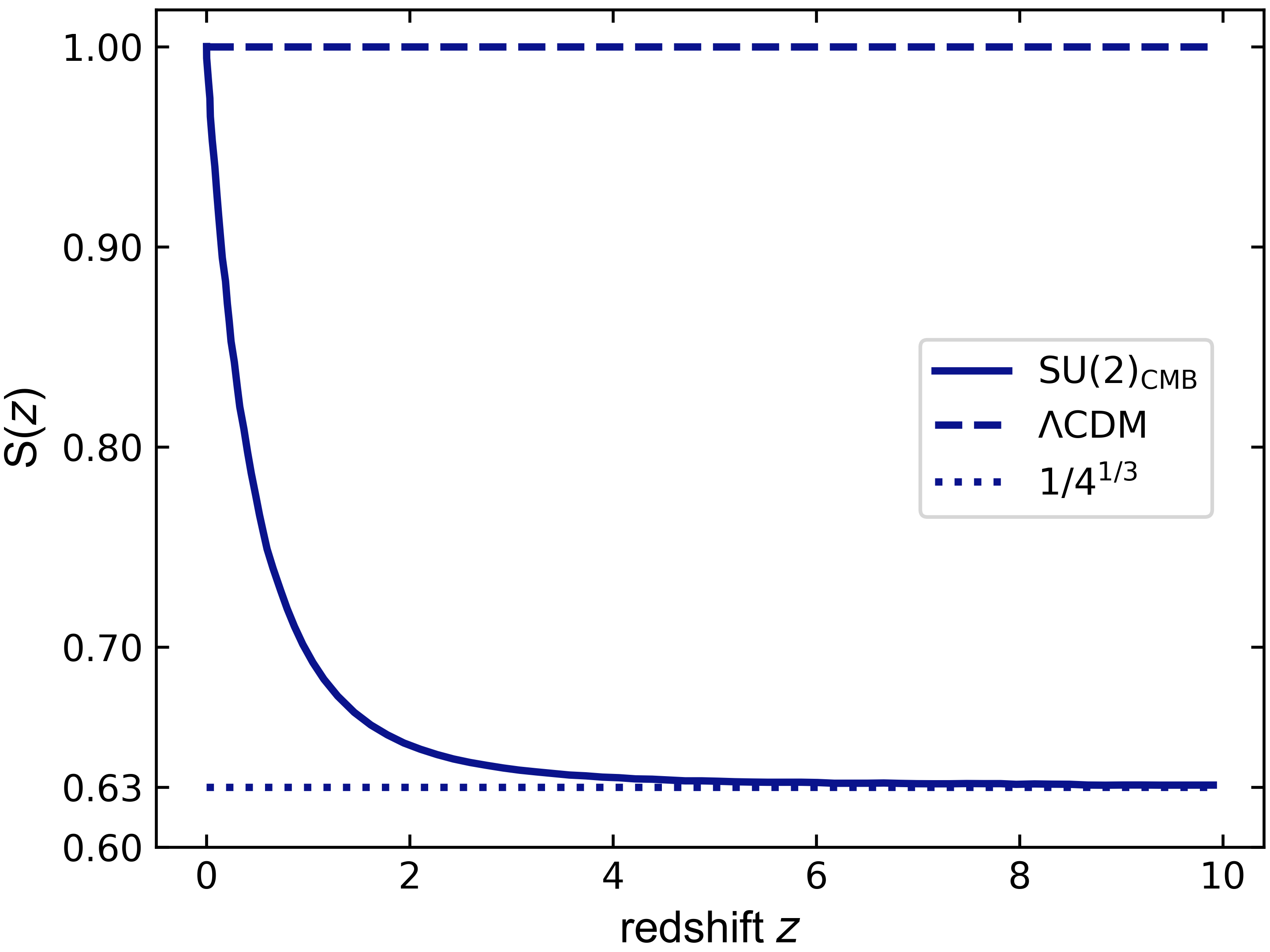}
\caption{\protect{\label{Fig-1}} Plot of function ${\cal S}(z)$ in Eq.\,(\ref{soltT0}). The curvature in ${\cal S}(z)$ at low $z$ indicates the breaking of conformal invariance in the deconfining SU(2) Yang--Mills plasma for $T\gtrsim T(z=0)$ with a rapid approach towards $\left({1}/{4}\right)^{1/3}$ as $z$ increases. The conventional $T$-$z$ relation of the CMB, as used in the cosmological standard model $\Lambda$CDM, is associated with the dashed line ${\cal S}(z)\equiv 1$. Figure adapted from \protect\cite{Hahn:2018dih}.}
\end{figure}

In deconfining SU(2)$_{\rm CMB}$ thermodynamics, asymptotic freedom \citep{Gross:1973id,Politzer:1973fx} occurs nonperturbatively for $T\gg T(z=0)$ \citep{2016tqym.book.....H}. The Stefan-Boltzmann limit is then well saturated, and therefore 
$s(T)$ is proportional to $T^3$. Moreover, at 
$T(z=0)$, due to a decoupling of massive vector modes, excitations represent a free photon gas. Therefore, $s(T(z=0))$ is proportional to $T^3(z=0)$. As a consequence, the ratio $s(T)/s(T(z=0))$ in Eq.\,(\ref{sol}) reads
\begin{align}
\label{ratentr}
\begin{split}
\frac{s(T)}{s(T(z=0))} &=\frac{g(T)}{g(T(z=0))} \left(\frac{T}{T(z=0)}\right)^3 \\
&=\left(\left(\frac{g(T)}{g(T(z=0))}\right)^{\frac{1}{3}}\frac{T}{T(z=0)}\right)^3\,,\ \ \ (T\gg T(z=0))\,,
\end{split}
\end{align}
where $g$ refers to the number of relativistic degrees of freedom at the respective 
temperatures. We have $g(T)=2\times 1+3\times 2=8$ (two photon polarizations plus three polarizations for each of the two vector modes) and $g(T(z=0))=2\times 1$ (two photon polarizations). Substituting this into Eq.\,(\ref{ratentr}), inserting the result into Eq.\,(\ref{sol}), and solving for $T$, we arrive at the high-temperature $T$-$z$ relation 
\eab
\label{solt>t0}
T&=&\left(\frac14\right)^{\frac13}(z+1)\,T(z=0)\nonumber\\ 
&\approx&0.629\,\,(z+1)\,T(z=0)\,,\ \ \  (T\gg T(z=0))\,.
\eae
Due to two vector modes of a finite, $T$ dependent mass contributing to $s(T)$ at low temperatures, the $T$-$z$ relation is modified as 
\begin{equation}
\label{soltT0}
T = {\cal S}(z)(z+1)\,T(z=0)\,, 
\ \ \  (T\ge T(z=0))\,,
\end{equation}
where the function ${\cal S}(z)$ is depicted in Fig.\,\ref{Fig-1}.
\begin{figure}
\centering
\includegraphics[width=\columnwidth]
{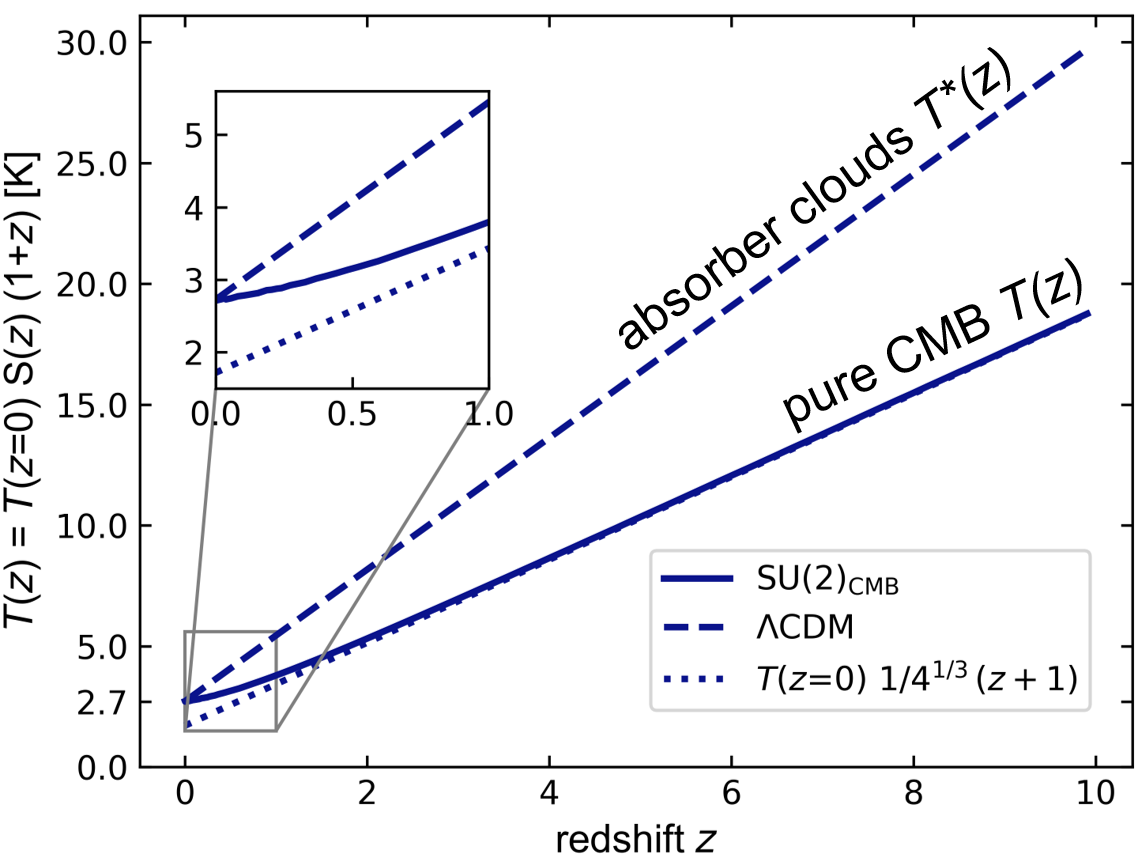}
\caption{\protect{\label{Fig-2}} 
Plot of the $T$-$z$ relation ${\cal S}(z) (1+z)$ in deconfining SU(2) Yang--Mills thermodynamics with $T(z=0)\equiv T_{\rm c}$. The conventional quantum U(1) $T$-$z$ relation of the CMB, employed in the cosmological standard model $\Lambda$CDM, is depicted by the dashed line.}
\end{figure}

In Section\,\ref{TRRCMB}, we reviewed why $T(z=0) \lesssim T_{\rm c}$. 
We now argue why $T(z=0)$ is excluded to be {\sl larger} than $T_{\rm c}$: There is another contribution to the excess in line temperatures at low frequencies from the fact that the frequency of waves (populating the deep Rayleigh-Jeans spectrum up to const/$T^2$) and the frequency of photons (starting to represent the spectrum for frequencies larger than const/$T^2$) redshift differently. On the other hand, baseline temperature $T(z)$ redshifts like the frequency of photons. That is, wavelike modes redshift as $\nu(z)=(1+z)\nu(z=0)$ while temperature (photon frequency) redshifts weaker than that like 
$T(z)=S(z)(z+1)T(z=0)$ with a numerically known function $S(z)<1$ of 
negative slope $\frac{dS(z)}{dz}\sim -1$ for $z\ll 1$, see Fig.\,\ref{Fig-1}. This {\sl also} contributes to an increase of line temperature at low frequencies compared to the Rayleigh-Jeans law since low frequencies are redshifted as usual but baseline temperature redshifts slower when lowering $z$ in the vicinity of $z=0$, see Fig.\,\ref{Fig-2} and Eq.\,(\ref{riseT}). 
Observationally \citep{Fixsen:2009xn}, the onset of this effect could be visible at $\nu\sim$\,1\,GHz which implies the wavelengths of low-frequency waves 
to be larger than 30\,cm, in turn, implying a critical temperature lower than 11.6\,K \citep{2016tqym.book.....H}. For the differential evolution of baseline temperature, we have 
\eqb
\label{riseT}
dT=T(z=0)\left(\frac{dS(z)}{dz}(1+z)+S(z)\right)\,dz\,.
\eqe
Since the present CMB’s line temperature rises steeply with a spectral index of $-2.6$ when lowering the frequency \citep{Fixsen:2009xn,Dowell:2018mdb}
the effect of Eq.\,(\ref{riseT}), which, modulo a mild stacking of low frequencies, is frequency independent 
for $\nu<\mbox{const}/T^2$, does not explain these large and variable line temperatures. Thus, we are again led to set $T(z=0)\sim T_{\rm c}$ to explain the observed steep rise in terms of wave evanescence (thermal Meissner mass). Therefore, the tuning $T(z=0)\sim T_{\rm c}$ is entirely explained by observation and does not require any ad hoc parameter coincidence.\\

Note that large and variable line temperatures cannot be explained in terms of the diffuse free-free emission facilitated by 
cosmological reionisation \citep{10.1093/mnras/stt2063,Oh_1999}. Interestingly, synchrotron radiation induced by weakly interacting massive particle (WIMPS) annihilations or decays in extra-galactic halos could match the low-frequency excess in CMB line temperature if a thermal annihilation cross section for light WIMPS is invoked \citep{PhysRevLett.107.271302}. Galactic radio emission is excluded as an explanation by the isotropy of the signal \citep{Seiffert_2011}. In \cite{PhysRevResearch013210} stochastic frequency diffusion is used to explain the low-frequency excess in the present CMB {(also dubbed `space roar´)} in terms of a primordial epoch of nonequilibrium conditions in the plasma. These conditions are modelled by a mild violation of the Einstein relation in the Kompaneets equation to allow for low-frequency localisation in the evolving photon distribution. The formation of the first generation of supermassive, cosmological black holes is speculated to explain {the} space roar in terms of synchrotron emission from the remnants \citep{10.1093/mnras/stu541}.\\

If the line-temperature excess can, indeed, be shown to persist to higher redshifts, including the onset of re-ionisation (cosmic dawn), then a potential explanation of the anomalously strong absorption of the redshifted
21-cm line by neutral hydrogen measured by the experiment to detect the global epoch of reionisation signature (EDGES) \citep{Bowman2018} is enabled. 
This would falsify our proposal of the present space roar to solely be a very-low redshift phenomenon due to an admixture of Gaussian distributed evanescent waves to the conventional low-frequency Rayleigh-Jeans CMB spectrum\footnote{This admixture would arise due to phase tunnelling occurring when supercooling the deconfining phase into the preconfining one in SU(2)$_{\rm CMB}$.}. However, the strong absorption of the redshifted
21-cm line can also be explained by dark-matter induced cooling of the absorbing cosmic gas without having to invoke excess intensity of the CMB at low frequencies throughout cosmic dawn \citep{Barkana2018}.

\subsection{Anisotropic photon emission by electrons or isotropic and homogeneous thermalisation\label{APE}}

In the framework of SU(2) Yang--Mills theory, 
why is it that spectral lines redshift according to the conventional 
$\nu$-$z$ relation $\nu(z)=(1+z)\nu(z=0)$ while the bulk of frequencies within the 
CMB follows a $\nu$-$z$ relation which associates with the $T$-$z$ relation 
of Section\,\ref{TRRCMB}?\\

The electron and its neutrino are modelled by a onefold-selfintersecting, figure-eight like center-vortex loop and a single center-vortex loop, respectively, see \cite{2017Entrp..19..575H,Hofmann:2020yhn} and \cite{2016tqym.book.....H}. These excitations are immersed into the 
confining ground-state of SU(2) Yang--Mills theory. A mass formula can be derived for the 
electron which equates the frequency of a breathing monopole \citep{PhysRevLett.92.151801,PhysRevLett.92.151802} 
(or the quantum selfenergy \citep{refId0,1964deBroglie}), contained 
within an extended ball-like blob associated with the region of vortex intersection, with the sum of the static monopole's rest mass 
and the energy content of deconfining SU(2) Yang--Mills 
thermodynamics of the blob (considering a mixing of two thermal gauge theories SU(2)$_{\rm CMB}$ and SU(2)$_{\rm e}$ at a temperature $T_{P=0}=1.18\,T_{c,\rm e}$ where the pressure vanishes \citep{futureAlphaPaper}). By invoking the value of the electron mass $m_{\rm e}=511\,$keV, this formula yields a value of the SU(2) Yang--Mills scale {$\Lambda_{\rm e}=3.62\,$keV} or a critical temperature {$T_{c,\rm e}=7.99\,$keV} for the deconfining-preconfining phase transition in SU(2)$_{\rm e}$ \citep{futureAlphaPaper}.\\

In addition, one obtains a blob radius $r_0\sim a_0$, where $a_0$ denotes the Bohr radius $a_0=0.592$ \AA. Also, the reduced Compton radius $r_c$, which roughly coincides with the core radius of the monopole $r_c$ \citep{PhysRevLett.92.151801,PhysRevLett.92.151802}, turns out to be $r_c\sim \alpha r_0$ where $\alpha\sim 1/137$ is the electromaganetic fine-structure constant. Modulo the electron's magnetic moment, carried by two closed vortex lines connecting to the blob, this matches with de Broglie's original interpretation of the electron \citep{refId0,1964deBroglie} and with the interpretation of the 
square of the wave function in wave mechanics \citep{https://doi.org/10.1002/andp.19263840404} as a probability density 
for locating a point particle \citep{https://doi.org/10.1007/BF01397477}. 
Namely, in its restframe, the electron represents an extended (spatially homogeneous) vibration induced by a charged monopole whose core size is negligible on the scale of the blob size and whose rate of jump-like location changes within the blob volume matches the vibration frequency $\nu_0$ ($m_0=h\nu_0$, $m_0$ the electron mass).\\  

If the global temperature of a photon gas is smaller than $T_{c,\rm e}$ then these 
photons must be thermalised with respect to SU(2)$_{\rm CMB}$ with 
$T_{c,\rm CMB}\sim 10^{-4}\,\mbox{eV}\ll 7.99\,\mbox{keV}\sim T_{c,\rm e}$. On the other hand, a directedly propagating electromagnetic field (a wave) represents a nonthermalised mode and thus cannot be subject to the gauge group 
SU(2)$_{\rm CMB}$ but rather is described by SU(2) Yang--Mills theories of much larger Yang--Mills scales \citep{2016tqym.book.....H}. The process of converting these isolated waves, emitted by the charge-carriers that do not penetrate into the volume bounded by a closed, emitting spatial surface, into a thermal photon 
gas contained within this volume hence proceeds by chopping their coherent intensity distribution into grainy and short-lived energy-momentum packets 
by an increasing homogenisation and isotropisation of energy transport as more and more differently directed waves of varying oscillation frequencies superposition away from the emitting surface. This very process of thermalisation, producing a photon gas with the temperature of the emitting surface, can effectively be understood as a rotation of SU(2) 
modes of theories with large Yang--Mills scales into those of SU(2)$_{\rm CMB}$.    

\subsection{Thermalisation dependent mixing of two SU(2) gauge theories}

For simplicity and due to its practical relevance\footnote{Charge carriers subject to SU(2) theories of 
larger Yang--Mills scales, represented by the charged leptons of the Standard Model $\mu^{\pm}$ and $\tau^{\pm}$, are 
instable due to weak decay and therefore do not qualify as material within the emitting surfaces of a blackbody cavity.} we 
consider the interplay of gauge groups SU(2)$_{\rm CMB}$ and SU(2)$_{\rm e}$.  The discussion in Section\,\ref{APE} can then be summarised as 
\eab
\label{rotationgf}
\bar{A}^{\rm CMB}_\mu&=&A^{\rm CMB}_\mu\,\cos\theta_W+A^{\rm e}_\mu\,\sin\theta_W\,,\nonumber\\ 
\bar{A}^{\rm e}_\mu&=&-A^{\rm CMB}_\mu\,\sin\theta_W+A^{\rm e}_\mu\,\cos\theta_W\,,
\eae
where ($\bar{A}^{\rm CMB}_\mu$,$\bar{A}^{\rm e}_\mu$) refers to the rotated state 
reached from the initial state ($A^{\rm CMB}$, $A^{\rm e}$)  
for the effective gauge fields in the deconfining phases of 
SU(2)$_{\rm CMB}$ and SU(2)$_{\rm e}$. The thermodynamically determined mixing angle $\theta_W$ turns out to be close to the electroweak mixing angle ($\theta_W=30.84^\circ$) if mixing within the interior of the blob -- representing the center of the region of selfintersection of the center-vortex loop -- is considered.  Within this central domain the mixed deconfining-phase pressure is demanded to vanish \citep{futureAlphaPaper}. Note that in case of infinite-volume thermodynamics at high temperatures (high-$z$ cosmology) such a stability constraint on a finite-volume region is irrelevant, and one has $\theta_W=45^\circ$.\\

In general, a change of the state of thermalisation induces a change of the rotation angle 
$\theta=\theta(\eta,T^*,T)$. In particular, for the interaction of the CMB with absorber clouds $\theta$ 
depends on the degree of thermalisation $\eta$ invoked by the initial states of electromagnetically interacting electrons of temperature $T^*$ within the cloud 
and the temperature $T$ of the CMB. 
Note that the thermalisation of these electronic states is influenced by these very lines dissipating directed background light. 
Via the degree of thermalisation $\eta$, the mixing angle $\theta$ also 
depends on the range of frequencies $\nu_l\le\nu\le\nu_u$ of wavelike modes in SU(2)$_{\rm CMB}$ and SU(2)$_{\rm e}$ which mediate the interactions between the 
electrons. Due to the present CMB exhibiting the largest low-frequency interval of excitations 
being associated with waves throughout its cosmological history, see Section\,\ref{TRRCMB}, 
and since the main frequency used for the extraction of $T(z=0)$ from background-source-absorber-cloud systems 
in the Milky Way \citep{1993ApJ...413L..67R} is 113.6\,GHz, which is to the left of the peak frequency $\nu= 160.4$\,GHz of the present CMB's blackbody spectrum, 
it is qualitatively understandable that the same temperature as from the 
CMB blackbody spectral fit in \citep{1994ApJ...420..439M} is observed for these systems. Due to the strong compression of the CMB wave spectrum with a factor $\propto T^{-2}$ it is also plausible that temperatures  
extracted from a background-source-absorber-cloud system at earlier epochs {\sl differ} from the associated CMB temperatures.\\

The quantitative computation of $\theta$ in a situation 
where a system of interacting (bound) electrons of a given number density, invoking a range of frequencies $\nu_l\le\nu\le\nu_u$ for these interactions, is immersed into the CMB, is a complex task 
which we hope to gain more insight about in the future.

\section{Summary and Outlook\label{SC}}

In this paper, we discussed two main approaches to extract the CMB temperature at finite redshifts: the 
analysis of absorption line profiles originating from gases of atoms, ions, or molecules within 
the line of sight of a broad-spectrum and bright background source, and the thermal Sunyaev-Zel'dovich effect (thSZ). In the literature, the 
assumption of a conventional frequency--redshift relation ($\nu$-$z$ relation) for CMB photons, considered to thermalise with relevant transitions in the cloud systems in the former case or to represent CMB spectral distortions in the latter situation, yields a conventional temperature--redshift relation ($T$-$z$ relation) for the CMB. We argued, based on the blackbody spectrum at all redshifts, that this is a consequence of thermodynamics. Whatever the assumed $\nu$-$z$ relation, observations necessarily produce the associated $T$-$z$ relation and vice versa. If the CMB is subject to an SU(2) rather than a U(1) quantum gauge principle, we reviewed how the corresponding $T$-$z$ relation changes. Consequently, the CMB's $\nu$-$z$ relation is changed. Finally, on a qualitative level, we provided reasons for which the temperature of a cloud of known redshift may differ from the temperature of a pure photon gas representing the CMB far away from the cloud. This is because thermalisation in the cloud, in addition the interaction of bound electrons with wavelike CMB disturbances, also proceeds via emissions and absorptions of wavelike modes by bound electrons. These modes, however, are subject to another (confining-phase) SU(2) Yang--Mills theory of a much higher critical temperature: SU(2)$_{\rm e}$.\\

The existence of SU(2)$_{\rm e}$ impacts Big Bang nucleosynthesis. Specifically, if {the electron is subject to an SU(2) gauge-theory model, involving the two factors SU(2)$_{\rm CMB}$ and SU(2)$_{\rm e}$ \citep{2017Entrp..19..575H,Hofmann:2020yhn,futureAlphaPaper} (see also \cite{MeinertHofmann2021}), then the Hagedorn temperature $T_H=6.66$ keV of SU(2)$_{\rm e}$ implies that the primordial Helium mass fraction of $Y=\frac14$ ($Y=\frac{2f_i}{1+f_i}$, where $f_i$ denotes the neutron-to-proton ratio at the onset of nucleosynthesis) is not induced by the nucleosynthesis of the light elements setting in at $T=65$ keV, subject to $f_i=\frac17$ (the freeze-out value $f=\frac15$ at $T\sim 800$ keV being reduced to $f_i=\frac17$ at $T=65$ keV due to neutron decay). Rather, nucleosynthesis would start at $T=65$ keV with $f_i=1$, implying a Helium mass fraction of $Y=1$ prior to the Hagedorn transition. The value of $Y$ would subsequently be reduced to $Y\sim \frac14$ through collective Helium photo-disintegration by gamma quanta that are released across the Hagedorn transition.} \\

Our conclusion regarding the extractions of the CMB temperature using the thSZ effect pursued in the literature is that they confirm the CMB blackbody spectrum at a finite redshift. However, under the assumed conventional $\nu$-$z$ relation, the result of the $T$-$z$ relation extraction is necessarily conventional as well. \\

The extraction of the CMB's $T$-$z$ relation from an assumed thermalisation within absorber clouds, which also uses a conventional $\nu$-$z$ relation for the relevant absorption lines, is questionable since the two systems, (i) a cloud immersed into the CMB and 
(ii) a pure CMB, exhibit different thermal degrees of freedom at a sufficiently high redshift: waves for (i) and photons for (ii).\\

One possibility for determining the effect of the $T$-$z$ relation subject to SU(2)$_{\rm CMB}$ is to study the flux of ultra-high-energy cosmic rays (UHECRs). In particular, there is a sensitive region below the ankle, that is, for $1\times 10^{18}\,\mbox{eV}\le E\le 6\times 10^{18}\,\mbox{eV}$. Due to the reduced CMB photon density at the same finite redshift, there is a higher flux of UHECRs under otherwise equal conditions for emission and propagation. Most prominently, the flux of protons is significantly increased in comparison to the use of the conventional $T$-$z$ relation when fitted to UHECR data 
\citep{Meinert:2023zhk2}.


\section{Acknowledgements}
This work was stimulated by an intense and fruitful exchange about a recent publication \citep{https://doi.org/10.1002/andp.202200517} between the two Referees and the current authors.\\

\noindent JM acknowledges insightful discussions with Karl-Heinz Kampert and Alexander Sandrock.\\

\noindent This work is supported by the Vector Foundation under grant number P2021-0102. 


\bibliographystyle{mnras}
\bibliography{BibSU2CMB}

\newpage

\label{lastpage}
\end{document}